\documentclass[aip,reprint,twocolumn,citeautoscript]{revtex4-1}
\usepackage[english]{babel}
\usepackage{amsmath}
\usepackage{diagbox}
\usepackage{hyperref}
\usepackage{cleveref}
\usepackage{float}
\usepackage{setspace}
\usepackage{graphicx}
\Crefname{equation}{Eq.}{Eqs.}
\creflabelformat{equation}{#2#1#3}
\crefrangelabelformat{equation}{#3#1#4 to~#5#2#6}

\renewcommand\[{\begin{equation}}
\renewcommand\]{\end{equation}}

\usepackage{xcolor}
\usepackage{hyperref} 
\hypersetup{ colorlinks, 
linkcolor={red!50!black}, 
citecolor={blue!80!black}, 
urlcolor={blue!80!black} }

\draft 

\begin{document}


\title{Device-scale modeling of valley photovoltaics} 



\author{Daixi Xia}
\affiliation{Department of Physics, University of Ottawa, Ottawa, Canada}

\author{Hassan Allami}
\affiliation{Department of Physics, University of Ottawa, Ottawa, Canada}

\author{Jacob J. Krich}
\affiliation{Department of Physics, University of Ottawa, Ottawa, Canada}
\affiliation{Nexus for Quantum Technologies, University of Ottawa, Ottawa, Canada}


\date{\today}

\begin{abstract}
We present a Poisson/drift-diffusion model that includes valley scattering effects for simulating valley photovoltaic devices. The valley photovoltaic concept is a novel implementation of a hot-carrier solar cell and leverages the valley scattering effect under large electric field to potentially achieve high voltage and high efficiency. Fabricated devices have shown S-shaped current-voltage curves, low fill factor, and thus low efficiency. We hence develop the first device model for valley photovoltaics. Our model includes electric-field-dependent valley scattering rates extracted from previous ensemble Monte Carlo simulations. We show that the condition of nonequilibrium carrier populations in the satellite valleys is not enough for valley photovoltaics to achieve high efficiency. We also show that increasing the built-in electric field of the valley-scattering region does not improve efficiency, contrary to previous suggestion.
\end{abstract}

\pacs{}

\maketitle 

\section{Introduction}

\begin{figure}
\begin{centering}
\includegraphics[width=0.5\textwidth]{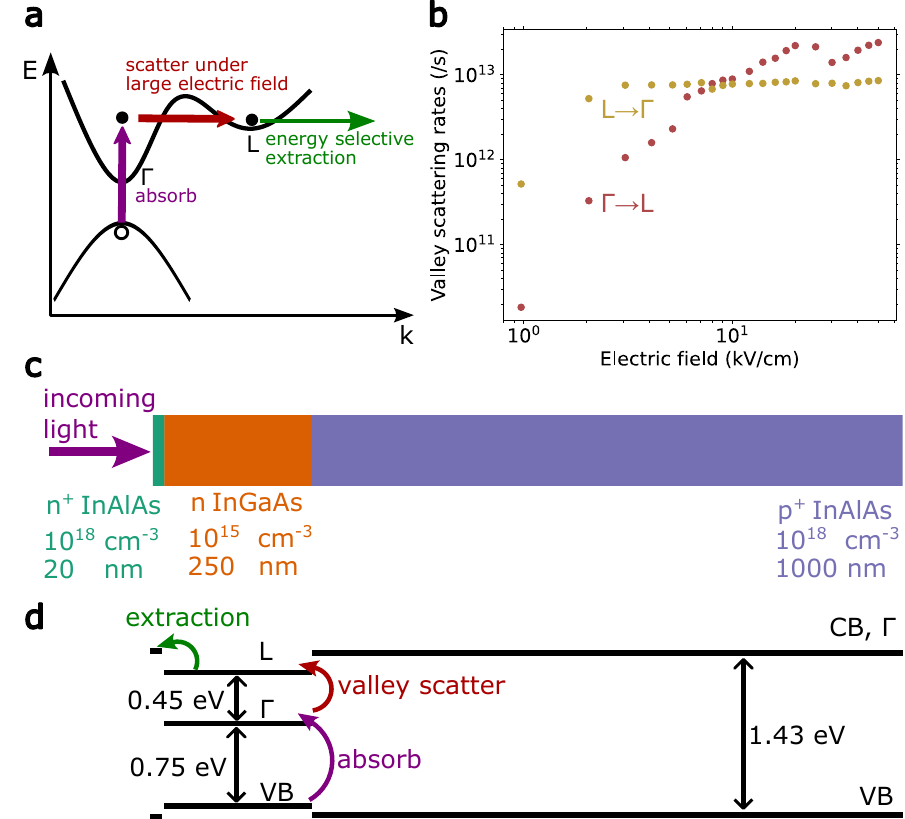}
\par\end{centering}
\caption{
\label{fig:concept}(a) Schematic of InGaAs band structure. VPV devices
are designed to facilitate net valley scattering from $\Gamma$ to
L valleys and aim to extract from L valleys. (b) 
Valley scattering rates extracted from EMC results\cite{ferry_search_2019}. $\Gamma$-to-L
rate exceeds L-to-$\Gamma$ at electric field larger than 10~kV/cm.
VPV device design (c) and band energy alignments (d) from Ref.~\onlinecite{esmaielpour_exploiting_2020}.
}
\end{figure}

Hot carrier solar cells have a high maximum theoretical efficiency
of 86\% under full concentration, significantly higher than the Shockley-Queisser
limit \cite{wurfel_particle_2005}. The high theoretical ceiling is
achieved by absorbing low-energy photons while reducing thermalization
loss in carriers generated by high-energy photons. A hot-carrier cell
must (1) sustain non-equilibrium carrier distributions and (2) have
energy-selective contacts \cite{green_third_2003}. The valley photovoltaic
(VPV) is a new concept for hot-carrier cells, illustrated in Fig.~\ref{fig:concept}(a),
where carriers produced in the $\Gamma$ valley in the Brillouin zone
of the conduction band scatter to satellite valleys. These metastable
valley populations can sustain non-equilibrium electron distributions
at a higher energy than the conduction band (CB) minimum. Extracting
electrons from these valleys results in potentially high operating
voltage. The valleys help simultaneously sustain high carrier energies
and provide an energy-selective contact. Ref.~\onlinecite{esmaielpour_exploiting_2020}
studied a VPV device sketched in Fig.~\ref{fig:concept}(c) with
n$^{+}$-In$_{0.52}$Al$_{0.48}$As, n-In$_{0.53}$Ga$_{0.47}$As and p$^{+}$-In$_{0.52}$Al$_{0.48}$As
on InP substrate. The design has a high built-in electric field in
InGaAs that is hypothesized to encourage valley scattering from the
$\Gamma$ valley to the L valley \cite{esmaielpour_exploiting_2020, ferry_search_2019, dorman_electric_2022}.
The InAlAs layers are designed to extract electrons from the L valley
of InGaAs, as illustrated in Fig.~\ref{fig:concept}(d). The device
in Ref.~\onlinecite{esmaielpour_exploiting_2020} shows distinct S-shaped
current-voltage (JV) curves and therefore low fill factor and efficiency,
as shown in Fig.~\ref{fig:1-14.5suns}. Ref.~\onlinecite{dorman_electric_2022}
suggested that a thinner InGaAs layer, hence larger built-in field,
produces more carriers in L through valley scattering (VS). Current
theoretical studies of VPV have used ensemble Monte Carlo (EMC) calculations
in homogenous InGaAs medium without the InAlAs layers and device details
\cite{ferry_search_2019}. To explain VPV devices, we need a computationally
efficient device model.

In this work, we develop a Poisson/drift-diffusion (PDD) model, with
separate populations in the $\Gamma$ and L valleys, each separately
quasi-thermalized at lattice temperature. We include valley scattering
as recombination and generation between $\Gamma$ and L populations
and use rates extracted from EMC results provided by David Ferry.
We incorporate those rates into the PDD model by introducing a quasi-electric
field that enables agreement with both EMC rates when carrier concentrations
are homogeneous and equilibrium detailed balance. In our model, we
include no genuine hot-carrier effect, as all carriers are thermalized
to the lattice temperature, but we consider carrier distributions
that are highly non-quasi-equilibrium when compared with a standard
two-band model of a semiconductor. We include the metastability of
the carriers in the satellite valleys as well as extraction from these
valleys. Our PDD model can qualitatively reproduce the experimental
JV curves, including the S shapes, even in absence of nonradiative
recombination processes. While S-shaped $J(V)$ curves are normally
associated with extraction barriers \cite{saive_s-shaped_2019}, our
simulation results show that the valley scattering process can also
cause S shapes. With our model, we study the effect of built-in electric
field by varying the thickness of the InGaAs region. We find that
increasing built-in electric field increases valley scattering from
$\Gamma$ to L in reverse bias but does not help with efficiency in
the forward bias where the device produces energy. Our model assumes that the hot CB carriers produced by valley scattering
can be described with separated $\Gamma$ and L valley populations
that are each quasi-equilibrated with the lattice. Any future efficient
VPV devices must violate this assumption; while we cannot simulate
such devices, deviations from predictions of our model can be used
to determine whether devices are operating in such a regime.

\section{Valley scattering effect and existing models}

Valley scattering is a type of electronic intraband transition caused
by phonon scattering that moves electrons between regions of the Brillouin
zone. The phonon momentum matches the momentum difference between
the initial and final valleys. By energy conservation, the energies
of the initial and final electronic states differ by the phonon energy.
With strong optical generation and valley scattering, the Fermi distribution is not sufficient to describe the CB population, even with an elevated carrier temperature \cite{ferry_search_2019}.

One can include valley scattering in some transport models,
such as the quantum Green's function formalism \cite{aeberhard_theory_2011}
and models based on solving the Boltzmann transport equation (BTE)
\cite{jacoboni_monte_1989,ferry_search_2019,blotekjaer_transport_1970,anile_two-valley_2002,mascali_hydrodynamical_2002}.
The quantum Green's function formalism has an advantage in simulating
nanostructures but can be computationally expensive for micrometer-scale
devices \cite{aeberhard_theory_2011}. BTE solves the semiclassical
electron dynamics and can be directly computed with ensemble Monte
Carlo (EMC) \cite{jacoboni_monte_1989}. EMC determines the energetic
distribution of electrons and does not assume a form for that distribution.
Ref.~\onlinecite{ferry_search_2019} studied the valley scattering effect
for photovoltaic purposes using an EMC model in a homogeneous medium.
The EMC formalism can include device transport. However, a device
EMC model requires larger computational resources than in a homogeneous
medium \cite{jacoboni_monte_1989}, so EMC models are generally used
for thin devices or in a hybrid model where EMC is applied in a small
portion of the device \cite{muralidharan_modeling_2022,baranowski_monte_2025}.

The BTE can also be computed indirectly by expanding in moments of
the population distribution. The infinite moment equations are truncated,
and a closure relation is asserted. PDD equations are the zeroth-
and first-moment equations of the BTE, where the electron distribution
function is Fermi-Dirac, while hydrodynamic (HD) models include a
small number of higher moments. The HD models were designed to study
Gunn oscillators and were also used for metal-oxide-semiconductor
field-effect transistors (MOSFET's) in the early 2000s. The HD models
allow for simulating energy transport independent of particle transport, allowing prediction of varying carrier temperature\cite{anile_two-valley_2002,blotekjaer_transport_1970}.
In the PDD case, the carrier population is determined completely by
the quasi-Fermi level, and the carrier temperature is equal
to the lattice temperature. We choose the PDD model for simulating
valley photovoltaics because it is computationally efficient when
including optical absorption and device transport, allows for a simple
implementation of valley scattering (detailed in Sec.~\ref{sec:PDD-VS}),
and is standard in photovoltaic calculations.

\section{PDD modeling of valley scattering\label{sec:PDD-VS}}

We develop a multi-band PDD model that includes both optical absorption
and valley scattering. We model valley scattering in a multi-band
PDD framework, using Simudo, which is a finite-element-based PDD device
model capable of simulating multiple bands with transport \cite{dumitrescu_simudo_2020}.
We treat the $\Gamma$ and the L valleys as separate bands, each with
its own carrier population with separate quasi-Fermi levels. 

In PDD, the carrier concentration $u_{k}$ in a band $k$ is related
to the current density $\boldsymbol{j}_{k}$ in that band by the drift-diffusion
and continuity equations,
\begin{align}
\boldsymbol{j}_{k} & =q\mu_{k}u_{k}\nabla w_{k}\label{eq:drift-diffusion}\\
\frac{\partial u_{k}}{\partial t} & =-s_{k}\frac{1}{q}\nabla\cdot\boldsymbol{j}_{k}+g_{k}^{\text{tot}},\nonumber 
\end{align}
where $s_{k}=-1$ for negative-charge carriers in a conduction band,
including $\Gamma$ and L valleys, $s_{k}=+1$ for holes in the valence
band, $\mu_{k}$ is the carrier mobility, $w_{k}$ is the quasi-Fermi
energy, and $q$ is the elementary charge. Eq.~\ref{eq:drift-diffusion}
contains both drift and diffusion terms and is valid for carrier populations
obeying Boltzmann and Fermi statistics \cite{poupaud_charge_1991}.
$g_{k}^{\text{tot}}$ is the total generation due to absorption and
recombination processes and can be separated into several terms, each
representing a different process,
\begin{equation}
g_{k}^{\text{tot}}=g_{k}^{\text{opt}}+g_{k}^{\text{rad}}+g_{k}^{\text{non-rad}}+g_{k}^{\text{VS}},
\end{equation}
where $g_{k}^{\text{opt}}$ is optical generation, $g_{k}^{\text{rad}}$
is radiative recombination, and $g_{k}^{\text{non-rad}}$ is non-radiative
recombination, such as Shockley-Read-Hall and Auger processes. Those
$g_{k}$ terms are standard in PDD device models. $g_{k}^{\text{VS}}$
is the net generation due to the valley-scattering process. Modeling
$g_{k}^{\text{VS}}$ is the focus of this study. In steady state,
$\frac{\partial u_{k}}{\partial t}=0$. The carrier distribution must
also obey Poisson's equation:
\[
\nabla\cdot\left(\epsilon\nabla\phi\right)=-\rho,
\]
where $\epsilon$ is the permittivity, $\phi$ is the electrostatic
potential, and $\rho$ is the charge density.

We assume parabolic dispersion for all bands, which gives the standard
connection between the quasi-Fermi levels $w_{k}$ and the carrier
concentrations,
\begin{equation}
u_{k}=\begin{cases}
\frac{1}{2\pi^{2}}\left(\frac{2m_{k}^{*}}{\hbar^{2}}\right)\int_{E_{k}}^{\infty}\frac{\left(E-E_{k}\right)^{3/2}\text{d}E}{e^{(E-w_{k}-q\phi)/k_{\text{B}}T}+1}, & \text{CB}\\
\frac{1}{2\pi^{2}}\left(\frac{2m_{k}^{*}}{\hbar^{2}}\right)\int_{-\infty}^{E_{k}}\frac{\left(E_{k}-E\right)^{3/2}\text{d}E}{e^{-(E-w_{k}-q\phi)/k_{\text{B}}T}+1}, & \text{VB},
\end{cases}\label{eq:fermi-population-full}
\end{equation}
where\textbf{ }$\hbar$ is the reduced Planck's constant,\textbf{
$m_{k}^{*}$ }is the effective mass, $E_{k}$ is band extremum, $\phi$
is the electrostatic potential, $k_{\text{B}}$ is the Boltzmann constant,
and $T=300$ K is the temperature. In bands where $s_{k}(w_{k}+q\phi-E_{k})\gg k_{\text{B}}T$,
Eq.~\ref{eq:fermi-population-full} reduces to the standard Boltzmann
result,
\begin{equation}
u_{k}=N_{k}e^{s_{k}\left(E_{k}-w_{k}-q\phi\right)/k_{\text{B}}T},\label{eq:boltz-population}
\end{equation}
where $N_{k}=2\left(\frac{m_{k}^{*}k_{\text{B}}T}{2\pi\hbar^{2}}\right)^{3/2}$
is the effective density of states. For degenerate bands where we
cannot take the Boltzmann approximation, numerically calculating $u_{k}(w_{k})$
and the inverse, $w_{k}(u_{k})$, is slow due to the Fermi integral.
We use the Joyce and Dixon approximation for parabolic bands, which
uses Lambert W function defined as $W(z)e^{W(z)}=z$ \cite{joyce_analytic_1977}.
Then, the population can be approximated as
\begin{equation}
u_{k}=\frac{N_{k}}{A_{1}}W\left[A_{1}e^{s_{k}\left(E_{k}-w_{k}-q\phi\right)/k_{\text{B}}T}\right],\label{eq:fermi-population}
\end{equation}
where $A_{1}=2^{-3/2}$. The closed-form inverse function is $w_{k}(u_{k})=k_{\text{B}}T\text{ln}\left[\frac{u_{k}}{N_{k}e^{s_{k}E_{k}/kT}}\right]$. 

We use a static absorption profile $g_{k}^{\text{opt}}$ with more
details described in Section~\ref{sec:Simulating-a-valley}. $g_{k}^{\text{rad}}$
is modeled with standard modified Planck emission spectrum assuming
energy-independent absorption coefficients, $\alpha_{i\rightarrow f}$,
where $i$ and $f$ are band indices. Standard Shockley-Read-Hall
(SRH) and Auger recombinations are also included in the $g_{k}^{\text{non-rad}}$
term. More details on Simudo can be seen in Ref.~\onlinecite{dumitrescu_simudo_2020}.
In this work, we only present results with $g_{k}^{\text{non-rad}}=0$,
because that is sufficient to capture the major qualitative features
of the existing candidate VPV devices.

We express valley scattering as generation and recombination in the
$\Gamma$ and L valleys, with electric-field-dependent rates $r_{k}$
extracted from the EMC results in Fig.~\ref{fig:concept}(b). The
total valley scattering rate is dependent on the populations in each
valley, 
\begin{equation}
g_{\text{L}}^{\text{VS}}=r_{\Gamma}u_{\Gamma}-r_{\text{L}}u_{\text{L}},\label{eq:g-vs}
\end{equation}
where $n_{\Gamma}$ and $n_{\text{L}}$ are the carrier concentrations
in the $\Gamma$ and L valleys, respectively. In equilibrium, $g_{\text{L}}^{\text{VS}}=0$.

Interpreting the EMC VS rates of Fig.~\ref{fig:concept}b naively
would present a contradiction at equilibrium. Specifically, for the
device shown in Fig.~\ref{fig:concept}(c), the built-in electric
field in the InGaAs region is $\simeq25$ kV/cm at equilibrium. The
EMC results in Fig.~\ref{fig:concept}(b) show that $r_{\Gamma}>r_{\text{L}}$
at 25~kV/cm. Since detailed balance requires that at equilibrium,
$g_{\text{L}}^{\text{VS}}=0$, we would conclude that $\frac{u_{\text{L}}^{\text{eq}}}{u_{\Gamma}^{\text{eq}}}>1$.
Here, $u_{\text{k}}^{\text{eq}}$ is the carrier population in band
$k$ at equilibrium and is obtained by setting $w_{k}=w^{\text{eq}}$
in Eq.~\ref{eq:boltz-population} or \ref{eq:fermi-population}.
However, in equilibrium, carrier populations of the valleys are decided
by the valley energy levels and the equilibrium Fermi level, which
is the same for both valleys' populations, so, in reality, $\frac{u_{\text{L}}^{\text{eq}}}{u_{\Gamma}^{\text{eq}}}\ll1$.
This contradiction arises because the EMC simulations are performed
in a spatially invariant sample. To resolve this apparent contradiction,
we introduce the quasi-electric field of a valley $k$ as 
\begin{equation}
\mathcal{E}_{k}=\frac{\nabla w_{k}}{q}.
\end{equation}
This replacement of the physical electric field with the gradient of a quasi-Fermi level is similar to the well-known formulation of band currents as $j_{k} = \mu_{k} u_{k} \nabla w_{k}$, which expresses drift and diffusion properties as a drift current with respect to the quasi-electric field \onlinecite{fichtner_semiconductor_1983}.
In the uniform-medium limit, the quasi-electric field is the physical
field: $\mathcal{E}_{k}=\nabla w_{k}/q=-\nabla\phi$. We can derive
this relation by noting that in uniform medium, carrier concentration
is constant, so from Eqs.~\ref{eq:boltz-population},\ref{eq:fermi-population},$\nabla\left[s_{k}\left(E_{k}-w_{k}-q\phi\right)/k_{\text{B}}T\right]=0$,
so in a uniform material, $\nabla E_{k}=0$ and $\nabla w_{k}=-q\nabla\phi$.
Therefore, our definition of $\mathcal{E}_{k}$ is consistent with
the uniform-medium EMC simulations. At equilibrium, $\mathcal{E}_{k}=0$,
which correctly gives $g_{\text{L}}^{\text{VS}}=0$. We reinterpret
the abscissa of Fig.~\ref{fig:concept}b as being $\mathcal{E}$
for each valley. Then, $r_{\Gamma}=r_{\Gamma}(\mathcal{E}_{\Gamma})$
and $r_{\text{L}}=r_{\text{L}}(\mathcal{E}_{\text{L}})$. We thus
incorporate valley scattering into the model, obeying both the EMC
limit and the carrier distribution at equilibrium. Unlike in EMC simulations,
we do not include the non-quasi-equilibrium carrier distribution or
elevated carrier temperature possible in real devices. In the steady-state
limit, the return to a Fermi-Dirac distribution is reasonable, but
the temperature is a bigger approximation. Reference~\onlinecite{esmaielpour_exploiting_2020}
reports measurements of the high-energy photoluminescence tail, which indicate elevated carrier temperatures in the $\Gamma$ valley. A hydrodynamic model would
be required to treat separate carrier temperatures in each valley
\cite{anile_two-valley_2002}.

The EMC data in Fig. \ref{fig:concept}b do not have data at equilibrium,
$\mathcal{E}=0$. Under the equilibrium condition, we know that $g_{\text{L}}^{\text{VS}}=0$,
so we have only one unknown equilibrium rate, since setting $r_{\text{L}}(0)$
determines $r_{\Gamma}(0)$. In principle, $r_{L}(0)$ is a calculable
physical quantity, but we do not currently have access to its value.
We take $r_{\text{L}}(0)$ as a tunable parameter. There are many
ways to connect the equilibrium scattering rates to the EMC data at
finite $\mathcal{E}$. We find that linear interpolation gives rise
to numerical instability. The lowest few $r_{L}$ and $r_{\Gamma}$
EMC points show a roughly quartic increase with $\mathcal{E}$. Therefore,
we opt to connect the equilibrium rates and the lowest EMC data points
with two-parameter quartic functions,
\begin{equation}
r_{k}(\mathcal{E}_{k})=a(\mathcal{E}_{k}-b)^{4},
\end{equation}
where the parameters $a$ and $b$ are determined by $r_{k}(0)$ and
the EMC data point with the lowest $\mathcal{E}$. 

The low-field extrapolation of $r_{\Gamma}^{\text{}}$ and $r_{\text{L}}^{\text{}}$
needs to obey equilibrium detailed balance in a device: $g_{\text{L}}^{\text{VS}}(\mathcal{E}_{k}=0,w_{k}=w^{\text{eq}})=0$.
We can satisfy this condition by setting $r_{\text{L}}(0)/r_{\Gamma}(0)=u_{\Gamma}(w^{\text{eq}})/u_{\text{L}}(w^{\text{eq}})$.
In this work, the $\Gamma$ valley carriers obey Fermi statistics,
Eq.~\ref{eq:fermi-population-full}, so the $w^{\text{eq}}$ dependence
is not eliminated in the ratio $u_{\Gamma}(w^{\text{eq}})/u_{\text{L}}(w^{\text{eq}})$.
Then, the equilibrium relation between $r_{\text{L}}(0)$ and $r_{\Gamma}(0)$
depends on $w^{\text{eq}}$:
\begin{equation}
r_{\Gamma}^{\text{}}(\mathcal{E}_{\Gamma}=0,w^{\text{eq}})=r_{\text{L}}^{\text{}}(0)\frac{u_{\text{L}}(w^{\text{eq}})}{u_{\Gamma}(w^{\text{eq}})}.\label{eq:zero_field_gamma}
\end{equation}
 At zero field but out of equilibrium, it is unphysical for $r_{\Gamma}^{\text{}}(\mathcal{E}_{\Gamma}=0)$
to still depend on $w^{\text{eq}}$. Therefore, in Eq.~\ref{eq:zero_field_gamma},
we use $w_{\Gamma}$ in place of $w^{\text{eq}}$. In our low-field
approximation, $r_{\Gamma}$ depends on both $\mathcal{E}_{\Gamma}$
and $w_{\Gamma}$.

An ideal VPV device has large $g_{\text{L}}^{\text{VS}}$, so the
device generally benefits from large $r_{\Gamma}$ and small $r_{\text{L}}$.
From Fig.~\ref{fig:concept}b, $r_{\Gamma}$ increases as $\mathcal{E}_{\Gamma}$
increases until $\mathcal{E}_{\Gamma}$ reaches around 20 kV/cm when
the increase of $r_{\Gamma}$ slows. For achieving small $r_{\text{L}}$,
we would need small $\mathcal{E}_{\text{L}}$, although $r_{\text{L}}$
is not sensitive to $\mathcal{E}_{\text{L}}$ when it exceeds 3 kV/cm.

\section{Simulating a valley photovoltaic device\label{sec:Simulating-a-valley}}

We study the structure shown in Fig.~\ref{fig:concept}(c), with
thicknesses and dopings as indicated. Table~\ref{tab:Simulation-parameters}
lists physical parameters used in our simulations. We use a static Beer-Lambert optical generation profile for each band, $g_{k}^{\text{opt}}(z)$, as a function of depth $z$ into the device. We calculate $g_{k}^{\text{opt}}(z)$ using the AM1.5G spectrum and energy-dependent absorption coefficients \cite{adachi_physical_1992, lumb_characterization_2014}. Here, we assume the InGaAs VB-to-L absorption is 1\% of the VB-to-CB absorption for the energies where it is energetically allowed.
Radiative recombination rates are calculated according to the van Roosbroeck--Shockley relation\cite{van_roosbroeck_photon-radiative_1954}.
All other material parameters are taken from~\onlinecite{palankovski_analysis_2004}.

\begin{table}
\caption{
\label{tab:Simulation-parameters}
Simulation parameters for In$_{0.52}$Al$_{0.48}$As and In$_{0.53}$Ga$_{0.47}$As lattice-matched to InP. 
All material parameters from~\onlinecite{palankovski_analysis_2004}.
}

\begin{centering}
\begin{tabular}{ll}
\hline 
Parameter & Value\tabularnewline
\hline 
\hline 
$\mu_{\text{L,InGaAs}}$ & 444 cm$^{2}$/V/s \tabularnewline
$\mu_{\Gamma\text{,InGaAs}}$ & $1.39\times10^{4}$ cm$^{2}$/V/s\tabularnewline
$\mu_{\text{VB,InGaAs}}$ & 490 cm$^{2}$/V/s\tabularnewline
$\mu_{\text{CB,InAlAs}}$ & 517 cm$^{2}$/V/s\tabularnewline
$\mu_{\text{VB,InAlAs}}$ & 136 cm$^{2}$/V/s\tabularnewline
$N_{\Gamma,\text{InGaAs}}$ & $2.10\times10^{17}$ cm$^{-3}$\tabularnewline
$N_{\text{L,InGaAs}}$ & $6.67\times10^{19}$ cm$^{-3}$\tabularnewline
$N_{\text{VB,InGaAs}}$ & $7.37\times10^{18}$ cm$^{-3}$\tabularnewline
$N_{\text{CB,InAlAs}}$ & $4.85\times10^{17}$ cm$^{-3}$\tabularnewline
$N_{\text{VB,InAlAs}}$ & $1.10\times10^{19}$ cm$^{-3}$\tabularnewline
$E_{\Gamma,\text{InGaAs}}$ & 0.72 eV\tabularnewline
$E_{\text{L,InGaAs}}$ & 1.25 eV\tabularnewline
$E_{\text{VB,InGaAs}}$ & 0 eV\tabularnewline
$E_{\text{CB,InAlAs}}$ & 1.31 eV\tabularnewline
$E_{\text{VB,InAlAs}}$ & -0.14 eV\tabularnewline
\hline 
\end{tabular}
\par\end{centering}
\end{table}

We simulate two carrier populations, in the valence and conduction bands,
in the InAlAs regions and three populations in the InGaAs region,
where $\Gamma$ and L valley populations are separate. In our model,
we use a thermionic boundary condition between the L valley of the
InGaAs region and the CB of InAlAs \cite{yang_numerical_1993}, and
we impose a non-conductive boundary condition for the $\Gamma$ valley
at the heterojunction interfaces. Therefore, carriers in the $\Gamma$
valley can only be extracted if they first scatter to the L valley.
These assumptions are optimistic for the voltage of the device, matching how the devices are proposed
to operate and are compatible with achieving high open-circuit voltage.
We have also considered carrier extraction directly from the $\Gamma$
valley (not shown here). We use Fermi statistics for the carriers
in the $\Gamma$ valley, while other bands have Boltzmann statistics.
The $\Gamma$ valley becomes highly degenerate close to the front
heterojunction, while other bands stay nondegenerate in the injection
levels relevant to this study. 

\begin{figure}
\includegraphics[width=0.5\textwidth]{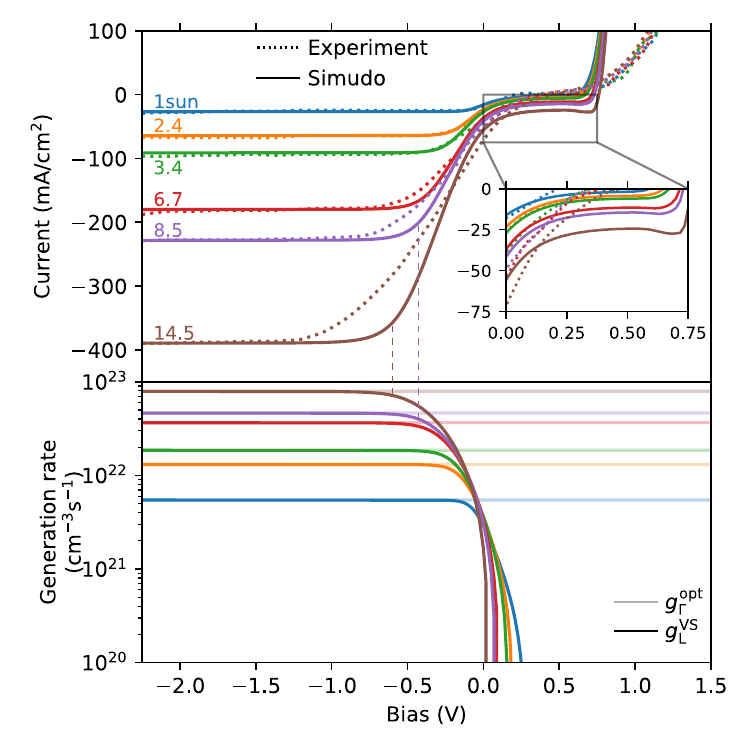}
\caption{\label{fig:1-14.5suns} (a) Current-voltage curves, simulated (solid)
and digitized from Ref.~\onlinecite{esmaielpour_exploiting_2020} (dotted).
The simulated currents are all multiplied by 1.09 to match the experimental
reverse saturation current at 14.5 suns. (b) Volume averaged valley
scattering generation rate to the L valley as a function of voltage
for the same illumination levels as in (a); the faded horizontal lines
indicate the volume averaged optical generation in the $\Gamma$ valley. The vertical dashed lines
show that the knees of the J(V) curves correspond to the voltages where the VS generation begins to fall below the optical generation.
In reverse bias, the device valley scatters to L all optically generated
electrons in $\Gamma$.}
\end{figure}

Our model qualitatively reproduces the experimental $J(V)$ curves
from~\onlinecite{esmaielpour_exploiting_2020}, as shown in Fig.~\ref{fig:1-14.5suns}a.
We observe in our simulations that the value of $r_{L}^{\text{}}(0)$
determines the amount of S shape in the $J(V)$ curves. In this work,
we choose $r_{L}^{\text{}}(0)=2\times10^{7}$~s$^{-1}$, which best
matches the knee in the S shape. Our model reproduces the shift of
the knee to more negative voltage with increased intensity. The reverse
saturation currents are determined by optical absorption. Our model
underestimates the reverse saturation currents by about 10\%. We adjust
all our simulated currents by a single factor of 1.09, such that our
reverse saturation current at 14.5 suns matches the experiment. With
this one adjustment factor, simulated currents at other intensities
match well with the experiments. As shown in the inset of Fig.~\ref{fig:1-14.5suns}a,
in the power-generating quadrant, we overestimate the current and
the open-circuit voltage. This current and voltage overestimate is
due to our model overestimating the absorption in the top InAlAs layer.
We have simulated versions (not shown) without optical absorption
in the top InAlAs, in which case our model shows better match to the
experiment of the $J(V)$ curves in the fourth quadrant. Our simulations use the nominal thicknesses of the device layers; if the as-grown devices had thinner top InAlAs layers than designed, the agreement with our simulations would be strong.
In forward bias, our model predicts a smaller ideality factor than the experimental
data, because we have only included radiative recombination, while
the device can have other recombination processes that exhibit larger
ideality factors. 

We show in Fig.~\ref{fig:1-14.5suns}b that the S shape in the $J(V)$
curves is caused by the reduction of $g_{\text{L}}^{\text{VS}}$ when
the voltage increases. In reverse bias, all optically generated electrons
in the $\Gamma$ valley are scattered to L and hence can be extracted.
However, as the voltage increases, $g_{\text{L}}^{\text{VS}}$ decreases
and eventually becomes negative. When $g_{\text{L}}^{\text{VS}}<0$,
the electrons are scattered from L to $\Gamma$, opposite to what
is desired in a high-efficiency VPV. The shoulders of $g_{\text{L}}^{\text{VS}}$
in Fig.~\ref{fig:1-14.5suns}b, where VS begins to decline, match
the onset of the S-shape in Fig.~\ref{fig:1-14.5suns}a, demonstrated
by the vertical dashed lines, with onset at lower $V$ with higher
intensity. 

\begin{figure}
\includegraphics[width=0.5\textwidth]{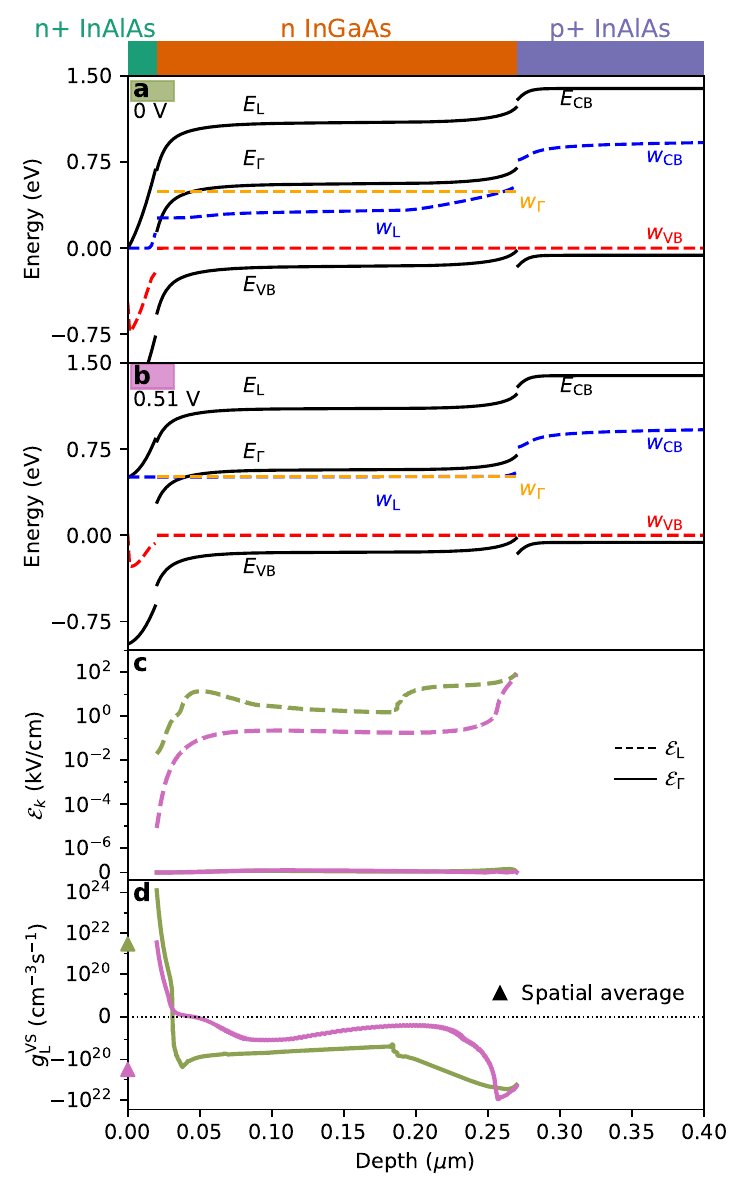}
\caption{\label{fig:one-sun-spatial} Simulated with one-sun illumination,
$r_{\text{L}}(0)=2\times10^{7}$ /s. Band diagrams at (a) short circuit
and (b) 0.51~V. (c) Quasi-electric field as a function of depth of
the device in the L (dashed) and $\Gamma$ (solid) valleys at short
circuit and 0.51~V. Colors correspond to the voltages in (a), (b).
(d) Valley scattering generation to L as a function of depth of the
device. The triangles indicate the spatial average.}
\end{figure}

We can understand why $g_{\text{L}}^{\text{VS}}$ decreases with voltage
by looking at band diagrams. Fig.~\ref{fig:one-sun-spatial}a,b shows
band diagrams under one sun at short circuit and 0.51 V. At short
circuit, the majority of the optically generated electrons are scattered
to L, while at 0.51 V, there is net transfer from L back to $\Gamma$,
reducing total current. From Eq.~\ref{eq:g-vs}, $g_{\text{L}}^{\text{VS}}$
depends on both $\mathcal{E}_{k}$ and $u_{k}$. We now take a closer
look at the spatially dependent $\mathcal{E}_{k}$. The band diagrams
in Fig.~\ref{fig:one-sun-spatial}a,b show that $w_{\Gamma}$ is
flat in both voltages, corresponding to near-zero $\mathcal{E}_{\Gamma}$,
plotted in Fig.~\ref{fig:one-sun-spatial}c. At short circuit, $\mathcal{E}_{\text{L}}$
is approximately 10 kV/cm in a large part of the InGaAs region and
is close to 100 kV/cm at the bottom. The resulting quasi-electric
field serves to increase the undesirable L-to-$\Gamma$ scattering.
At 0.51~V, $\mathcal{E}_{\text{L}}$ is smaller than 1 kV/cm except
towards the bottom, where $\mathcal{E}_{\text{L}}$ has a similar
value to the short-circuit case. Now we move our examination from
$\mathcal{E}_{k}$ to $u_{k}$. As seen in the band diagrams in Fig.~\ref{fig:one-sun-spatial}a,b,
the populations are largest at the front of the InGaAs region, due
to strong optical absorption at the first 20 nm. In particular, $u_{\Gamma}>u_{\text{L}}$
because the direct-gap VB-to-$\Gamma$ transition has a much larger
absorption coefficient. Therefore, despite ${\cal E}_{\text{L}}>{\cal E}_{\Gamma}$,
the effect of carrier population outweighs the effect of quasi-electric
fields. As shown in Fig.~\ref{fig:one-sun-spatial}d, the resulting
$g_{\text{L}}^{\text{VS}}$ is then largest at the front of the InGaAs.
In forward bias, at $0.51$~V, the front $g_{\text{L}}^{\text{VS}}$
is smaller than that at short circuit. Therefore, the volume-total
$g_{\text{L}}^{\text{VS}}$ at $0.51$~V is negative.

Efficient VPV requires achieving a voltage larger than the InGaAs
$\Gamma$-valley bandgap $E_{\text{g}}^{\Gamma}$, but both experimental
and simulated devices have $V_{\text{oc}}<E_{\text{g}}^{\Gamma}$.
In our model, the small $V_{\text{oc}}$ occurs because the net VS
generation is dominated by the zero-field VS rates. If we consider
$w_{\text{L}}$ to be spatially constant, as is approximately true
in forward bias (Fig.~\ref{fig:one-sun-spatial}b), then $V_{\text{oc}}$
is limited by $w_{\text{L}}-w_{\text{VB}}$, with $w_{\text{VB}}$
evaluated at the p-type contact, at the back of the device. When $\mathcal{E}_{k}\ll1$~kV/cm,
from Eqs.~\ref{eq:g-vs} and \ref{eq:zero_field_gamma}, the net
scattering rate is:
\begin{equation}
g_{L}^{\text{VS}}(\mathcal{E}_{\Gamma},\mathcal{E}_{L}=0)=n_{L}r_{L}^{\text{}}(0)\left[\text{exp}\left(\frac{w_{\Gamma}-w_{L}}{k_{\text{B}}T}\right)-1\right].
\end{equation}
From this form, we can see that $g_{\text{L}}^{VS}$ is negative when
$w_{L}>w_{\Gamma}$. Therefore, to efficiently scatter $\Gamma$ valley
electrons to L, $w_{\text{L}}$ needs to be less than $w_{\Gamma}$.
Since radiative recombination forces $w_{\Gamma}-w_{\text{VB}}<E_{g}^{\Gamma}$,
if there is positive current from $\Gamma$ to L, then $w_{\text{L}}-w_{\text{VB}}<w_{\Gamma}-w_{\text{VB}}<E_{g}^{\Gamma}$.
Therefore, the device's $V_{\text{oc}}$, bounded by $w_{\text{L}}-w_{\text{VB}}$,
cannot exceed $E_{\text{g}}^{\Gamma}$. This constraint rules out
the possibility of high-efficiency VPV devices, which must be able
to obtain higher voltages.

The above simulations agree with the experimental results that the
studied InGaAs VPV device does not produce high efficiency. We now
consider what happens to VS efficiency if we allow the material properties
to change, as in a hypothetical material with stronger VS processes.
In Sec.~\ref{sec:thicks} we return to InGaAs and consider the effects
of changing the thickness --- and thus the equilibrium electric field
--- of the InGaAs region.

\section{Effects of equilibrium valley scattering rates on S shape J(V)\label{sec:eq_rate}}

\begin{figure}
\begin{raggedright}
\includegraphics[width=0.5\textwidth]{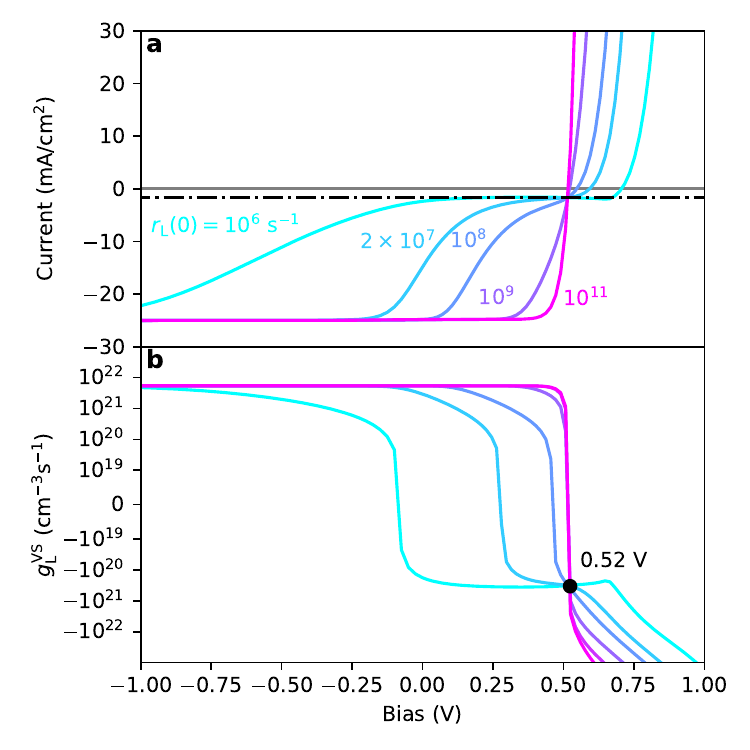}
\par\end{raggedright}
\caption{\label{fig:eq_rates_jv_vs_gen}(a) One-sun $J(V)$ curves simulated
with different values of $r_{\text{L}}^{\text{}}(0)$. The $J(V)$ matches experiments the best with $r_{\text{L}}^{\text{}}(0) = 2\times10^7$. The dot-dash
line is the photocurrent in the top InAlAs region. (b) Spatially averaged
$g_{\text{L}}^{\text{VS}}$ as a function of voltage for various $r_{\text{L}}^{\text{}}(0)$
values.}
\end{figure}

We consider a range of hypothetical materials by considering the effects
of changing $r_{\text{L}}(0)$ and show that the S shape in the $J(V)$
can be eliminated by increasing $r_{\text{L}}(0)$, but $V_{\text{oc}}$
is still limited by the InGaAs bandgap. Fig.~\ref{fig:eq_rates_jv_vs_gen}a
shows the one-sun $J(V)$ curves simulated for the same device structure
shown in Fig.\ref{fig:concept}c. Smaller $r_{L}^{\text{}}(0)$ values
result in smaller fill factors, i.e., more ``S'' shaped $J(V)$.
For $r_{\text{L}}(0)\gtrsim10^{9}$ s~$^{-1}$, the $J(V)$ becomes
diode like. Although larger $r_{L}^{\text{}}(0)$ increases fill factor,
the $V_{\text{oc}}$ decreases asymptotically to $\simeq0.52$~V.

The S-shaped JV and low fill factor are due to insufficient net valley
scattering from the $\Gamma$ to the L valley. Fig. ~\ref{fig:eq_rates_jv_vs_gen}(b)
shows volume-averaged $g_{\text{L}}^{\text{VS}}$ as a function of
voltage. In reverse bias, valley scattering can bring all optically
generated electrons in $\Gamma$ to L for all $r_{\text{L}}^{\text{}}(0)$
values except the lowest case, but the net VS generation rate decreases
as the voltage increases. With larger $r_{\text{L}}^{\text{}}(0)$,
VS generation decreases more slowly as the voltage becomes positive.
However the trend reverses as V approaches 0.52~V, where larger $r_{\text{L}}^{\text{}}(0)$
results in more negative VS generation, opposing photocurrent.

\begin{figure}
\includegraphics[width=0.5\textwidth]{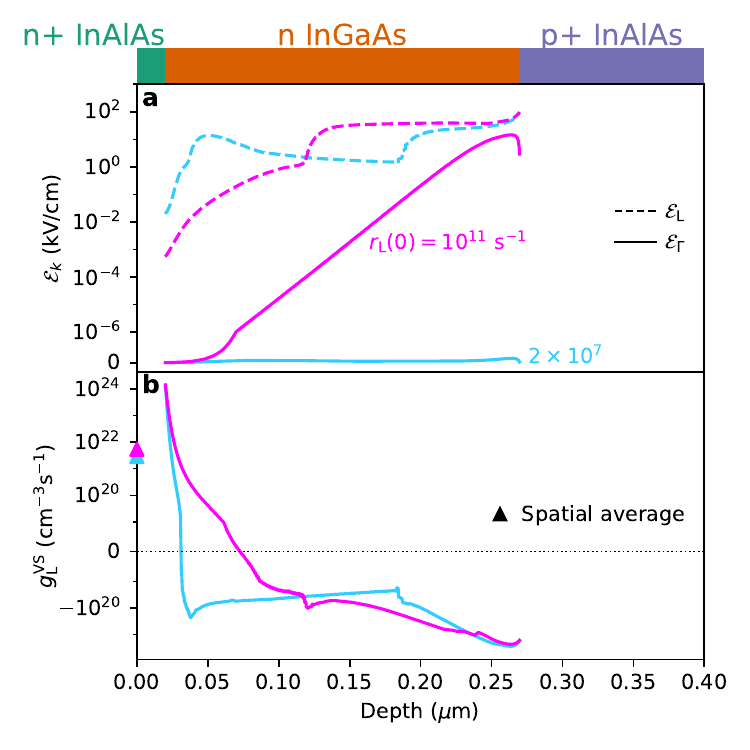}

\caption{\label{fig:eq_rates_spatial}Under one-sun illumination, 0~V, (a)
$\mathcal{E}_{k}$ and (b) $g_{\text{L}}^{\text{VS}}$ as a function
of device depth, for the two choices of $r_{\text{L}}(0)$ in Fig.~\ref{fig:eq_rates_jv_vs_gen}. $r_{\text{L}}^{\text{}}(0) = 2\times10^7$ gives the best match of J(V) to experiments.
The volume averaged $g_{\Gamma\rightarrow\text{L}}^{0}$ are indicated
as triangles on the y-axis.}
\end{figure}

Under our model, for a VPV device to achieve high efficiency, we need
to have a large $\mathcal{E}_{\Gamma}$, which could permit $g_{\text{L}}^{\text{VS}}$
to be larger while still having $w_{\text{L}}>w_{\Gamma}$, which
would enable increased $V_{\text{oc}}$. In Fig \ref{fig:eq_rates_spatial}a,
$\mathcal{E}_{\Gamma}$ becomes large near the bottom of the InGaAs
region for $r_{L}^{\text{}}(0)=10^{11}$~s$^{-1}$. Fig~\ref{fig:eq_rates_spatial}b,
which plots the net VS generation to L, shows that despite this large
$\mathcal{E}_{\Gamma}$, $g_{\text{L}}^{\text{VS}}$ is still small at the bottom,
since the $n_{\Gamma}$ in this region is small. Hence, the advantage
of large $\mathcal{E}_{\Gamma}$ does not show in the net current.
Future designs of VPV can focus on creating large $\mathcal{E}_{\Gamma}$
on the top portion of InGaAs, where $u_{\Gamma}$ is large.

\section{Large built-in field does not improve efficiency\label{sec:thicks}}

Ref.~\onlinecite{dorman_electric_2022} proposed that a thinner InGaAs
region, hence with larger built-in electric field, results in more
valley scattering to L, which is beneficial to VPV devices. In this
section, we test this hypothesis with our model and show that larger
built-in field does not facilitate more generation to L through valley
scattering in forward bias and therefore does not improve efficiency.

\begin{figure}
\centering \includegraphics[width=0.5\textwidth]{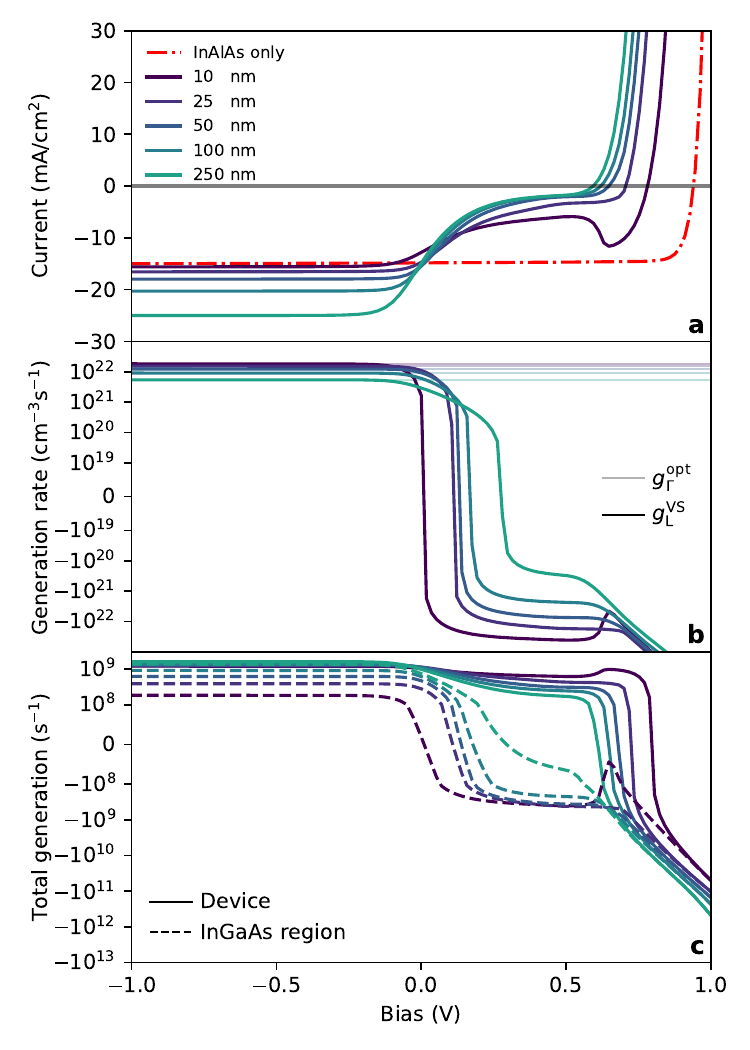}
\caption{\label{fig:thicks} (a) Current-voltage curves at one sun, with various
InGaAs layer thicknesses and for a reference device without the InGaAs;
(b) volume integrated $g_{\text{L}}^{\text{VS}}$ and $g_{\text{L}}^{\text{opt}}$;
(c) net generation to InAlAs CB and InGaAs L valley from optical and
valley scattering in all regions (solid) and in the InGaAs layer only
(dashed).}
\end{figure}

In our simulations, we vary the thickness of InGaAs from 10 to 250~nm.
The structures with thinner InGaAs regions have larger built-in fields
than those with thicker InGaAs. Fig.~\ref{fig:thicks}a
shows increasing fill factor with thinner InGaAs, despite reduced reverse
saturation currents as the InGaAs absorber becomes thin. The increased
fill factor is consistent with that reported in Ref.~\onlinecite{dorman_electric_2022}.
Our model shows that the cause of this phenomenon is not as proposed
in Ref.~\onlinecite{dorman_electric_2022}.

The increased fill factor comes solely from the increased absorption
in the bottom InAlAs region, due to higher transparency as the InGaAs
layer is made thinner. Fig.~\ref{fig:thicks}b shows $g_{\text{L}}^{\text{VS}}$
and $g_{\text{L}}^{\text{opt}}$, averaged over volume. In structures
with thinner InGaAs, the net valley scattering rate is larger in reverse
bias, due to larger per-volume optical absorption. We plot $g_{\text{L}}^{\text{VS}}$
and $g_{\text{L}}^{\text{opt}}$ as functions of voltage in Fig.~\ref{fig:thicks}b.
In the reverse saturation region, $g_{\text{L}}^{\text{VS}}=g_{\text{L}}^{\text{opt}}$
for the devices of all InGaAs-layer thickness. No matter how large
the built-in field, all structures are capable of valley scattering
to L all of the optically generated electrons in $\Gamma$. However,
$g_{\text{L}}^{\text{VS}}$ decreases rapidly around $V=0$. In forward
bias, near maximum power point (mpp), $g_{\text{L}}^{\text{VS}}<0$
for all structures, hurting the overall current. Although the thinnest
InGaAs has larger $V_{\text{oc}}$ and larger current at mpp as shown
in Fig.~\ref{fig:thicks}a, the thinnest InGaAs actually has its
forward-bias current most strongly reduced by valley scattering. In
Fig.~\ref{fig:thicks}b, $g_{\text{L}}^{\text{VS}}$ is more negative
with thinner InGaAs. In the thinnest case with 10 nm InGaAs, around
0.6~V, the device has a nonmonotonic $J(V)$ and $g_{\text{L}}^{\text{VS}}$.
However, $g_{\text{L}}^{\text{VS}}$ is still negative when the current
magnitude increases, so the current and the efficiency are still hurt
by valley scattering in forward bias. Eliminating the InGaAs layer
altogether results in the largest fill factor, largest $V_{\text{oc}}$,
and highest efficiency, as shown with the red $J(V)$ curve in Fig.~\ref{fig:thicks}a.

To see the cause for the increased fill factor with thinner InGaAs,
we plot $g_{\text{L}}^{\text{tot}}$ in the InGaAs region, including
valley scattering, optical generation, and radiative recombination,
as dashed lines in Fig.~\ref{fig:thicks}c. The solid lines in Fig.~\ref{fig:thicks}c
are sums of $g_{\text{L}}^{\text{tot}}$ in InGaAs and $g_{\text{CB}}^{\text{tot}}$
in InAlAs. The difference between the solid and the dashed lines is
the optical generation, net of radiative recombination, in the InAlAs
regions. With thinner InGaAs, more light reaches the bottom InAlAs
layer. We see that the optical generation in the bottom InAlAs explains
the phenomenon where thinner InGaAs region increases current and fill
factor. 

\section{Discussion}

We have performed device-level simulations of proposed VPV devices
using a PDD model. These simulations include quasi-equilibrium carriers
in each of the bands so do not include genuine hot-carrier effects
beyond quasi-equilibrium. We nonetheless find compelling qualitative
agreement with experimental J(V) curves. Within this model, it does
not appear that VPV devices can achieve high efficiency. The large
reverse-saturation current is consistent with the small $\Gamma$
bandgap in InGaAs, which allows the device to absorb a larger portion
of the solar spectrum than the larger-bandgap InAlAs. In principle,
if it could maintain separated quasi-Fermi levels of $\Gamma$ and
L while extracting only from L, the device's voltage would be capable
of exceeding the small $\Gamma$ bandgap. We can draw a loose analogy
between VPV and intermediate-band solar cells (IBSC), where $\Gamma$
valley corresponds to intermediate band, and L valley corresponds
to IBSC's CB. In IBSC, an electron in IB absorbs a photon and transfers
to the CB, while in VPV, an electron in $\Gamma$ valley scatters
to L. In an intermediate-band solar cell (IBSC), the separated quasi-Fermi
levels of the intermediate band (IB) and CB allow for a large voltage
that is limited by CB-VB bandgap while at the same time leveraging
the IB for absorbing sub-gap photons and increasing photocurrent.
Unlike in IBSC, we only observe the increased photocurrent due to
VS in reverse bias, and in forward bias, the VS process limits the
voltage to the $\Gamma$ bandgap, instead of the larger L gap. Our
model does not include carrier-temperature effects. Deviations from the predictions
of our model could be used to prove the existence of non-quasi-equilibrium
effects in experiments and whether high efficiencies are possible. Our results show that a
high-efficiency VPV device must have not only nonequilibrium populations in the satellite valleys, but there must also be elevated carrier temperatures. This conclusion agrees with the original VPV proposal, but existing electrical measurements on current devices are explainable without contribution from any such high-carrier-temperature populations.

\section{Acknowledgements}

We thank Ian Sellers, David Ferry, Ned Ekins-Daukes, and their groups
for valuable discussions.

We acknowledge NSERC CREATE TOP-SET (Award 497981) for funding.

\bibliography{references}

\end{document}